\documentclass[pra,twocolumn,nofootinbib,floatfix]{revtex4}

\usepackage[usenames]{color}
\usepackage{amsmath}
\usepackage{amssymb}
\usepackage{float}

\usepackage{graphicx}
\usepackage{xspace}

\newcommand{\ket}[1]{\ensuremath{\left|#1\right\rangle}}
\newcommand{\nbari}{\bar n_i}
\newcommand{\nbaridet}{\bar n_{i,\rm det}}

\newcommand{\mul}{\mu_{\mathrm{loc}}}
\long\def\symbolfootnote[#1]#2{\begingroup%
\def\thefootnote{\fnsymbol{footnote}}\footnotetext[#1]{#2}\endgroup}

\begin{document}

\title{Single-Atom Resolved Fluorescence Imaging of an Atomic Mott Insulator}

%% Notice placement of commas and superscripts and use of &
%% in the author list
\author{Jacob F. Sherson$^{1*\dag}$}
\author{Christof Weitenberg$^{1*}$}
\author{Manuel Endres$^{1}$}
\author{Marc Cheneau$^{1}$}
\author{Immanuel Bloch$^{1,2}$}
\author{Stefan Kuhr$^{1\ddag}$}

\date{29 July 2010}

\affiliation{
   $^1$Max-Planck-Institut f\"ur Quantenoptik, Hans-Kopfermann-Str. 1, 85748 Garching,
   Germany\\
   $^2$Ludwig-Maximilians-Universit\"at, Schellingstr. 4/II, 80799 M\"unchen, Germany
}

\begin{abstract}
The reliable detection of single quantum particles has
revolutionized the field of quantum optics and quantum information
processing. For several years, researchers have aspired to extend
such detection possibilities to larger scale strongly correlated
quantum systems, in order to record in-situ images of a quantum
fluid in which each underlying quantum particle is detected. Here we
report on fluorescence imaging of strongly interacting bosonic Mott
insulators in an optical lattice with single-atom and single-site
resolution. From our images, we fully reconstruct the atom
distribution on the lattice and identify individual excitations with
high fidelity. A comparison of the radial density and variance
distributions with theory provides a precise in-situ temperature and
entropy measurement from single images. We observe Mott-insulating
plateaus with near zero entropy and clearly resolve the high entropy
rings separating them although their width is of the order of only a
single lattice site. Furthermore, we show how a Mott insulator melts
for increasing temperatures due to a proliferation of local defects.
Our experiments open a new avenue for the manipulation and analysis
of strongly interacting quantum gases on a lattice, as well as for
quantum information processing with ultracold atoms. Using the high
spatial resolution, it is now possible to directly address
individual lattice sites. One could, e.g., introduce local
perturbations or access regions of high entropy, a crucial
requirement for the implementation of novel cooling schemes for
atoms on a lattice.
\end{abstract}

\maketitle
% INTRODUCTION

\symbolfootnote[1]{These authors contributed equally to this work.}
\symbolfootnote[2]{present address: Department of Physics and
Astronomy, University of Aarhus, DK-8000 Aarhus C, Denmark.}

\symbolfootnote[3]{Electronic address: \url{stefan.kuhr@mpq.mpg.de}}
\begin{figure}[]
    \begin{center}
        \includegraphics[width=\columnwidth]{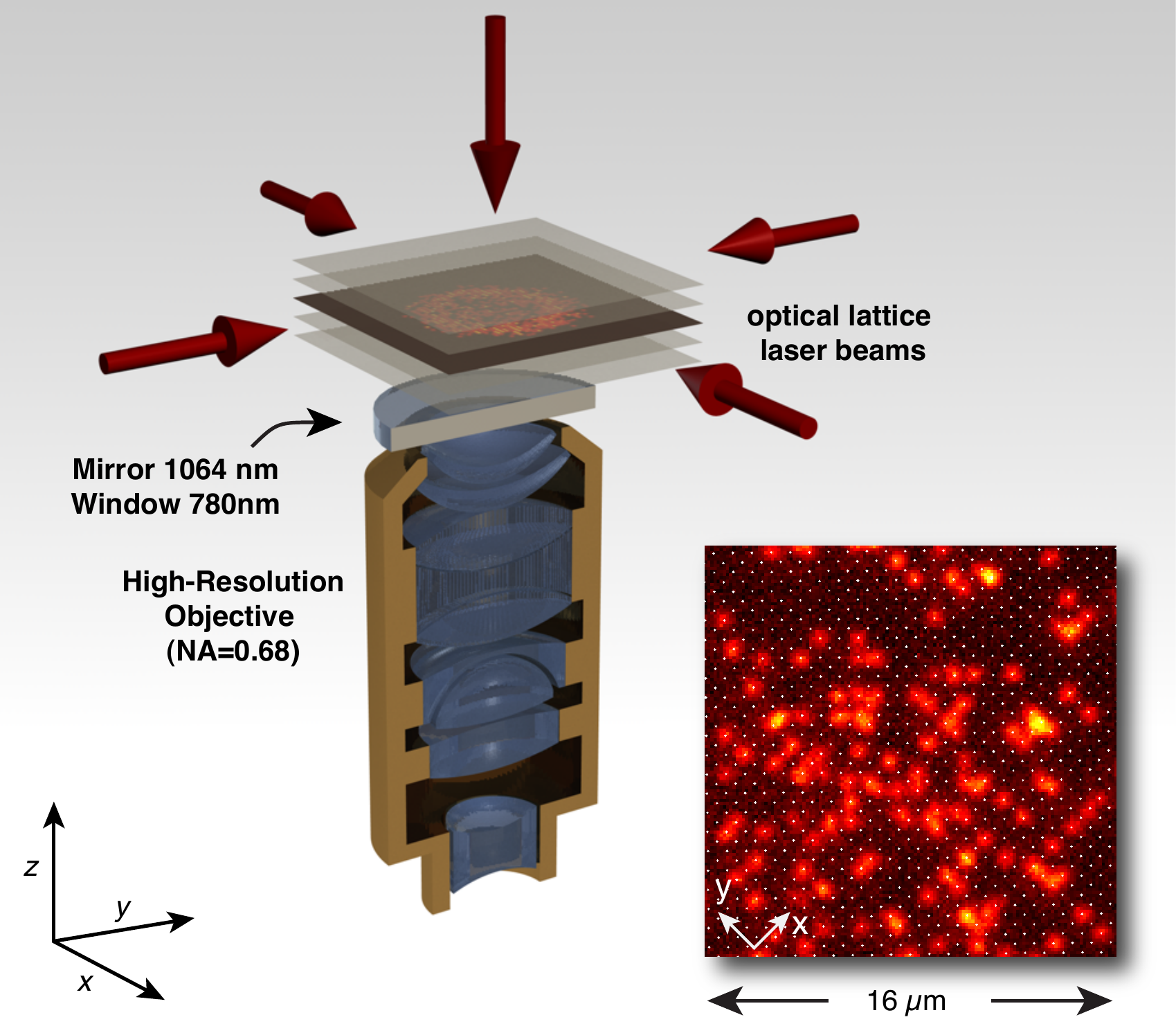}
    \end{center}
    \caption{{\bf Experimental Setup.} Two-dimensional bosonic quantum gases are prepared in a single 2D plane of an optical standing wave along the $z-$direction, which is created by retroreflecting a laser beam ($\lambda=1064$\,nm) on the coated vacuum window. Additional lattice beams along the $x$- and $y$-directions are used to bring the system
    into the strongly correlated regime of a Mott insulator. The atoms are detected using fluorescence imaging via a high resolution microscope objective. Fluorescence of the atoms was induced by illuminating the quantum gas with an optical molasses that simultaneously laser cools the atoms. The inset shows a section from a fluorescence picture of a dilute thermal cloud (points mark the lattice sites). \label{fig:schematic}}
\end{figure}
\begin{figure*}[!ht]
    \begin{center}
        \includegraphics[angle=0,width=2.0\columnwidth]{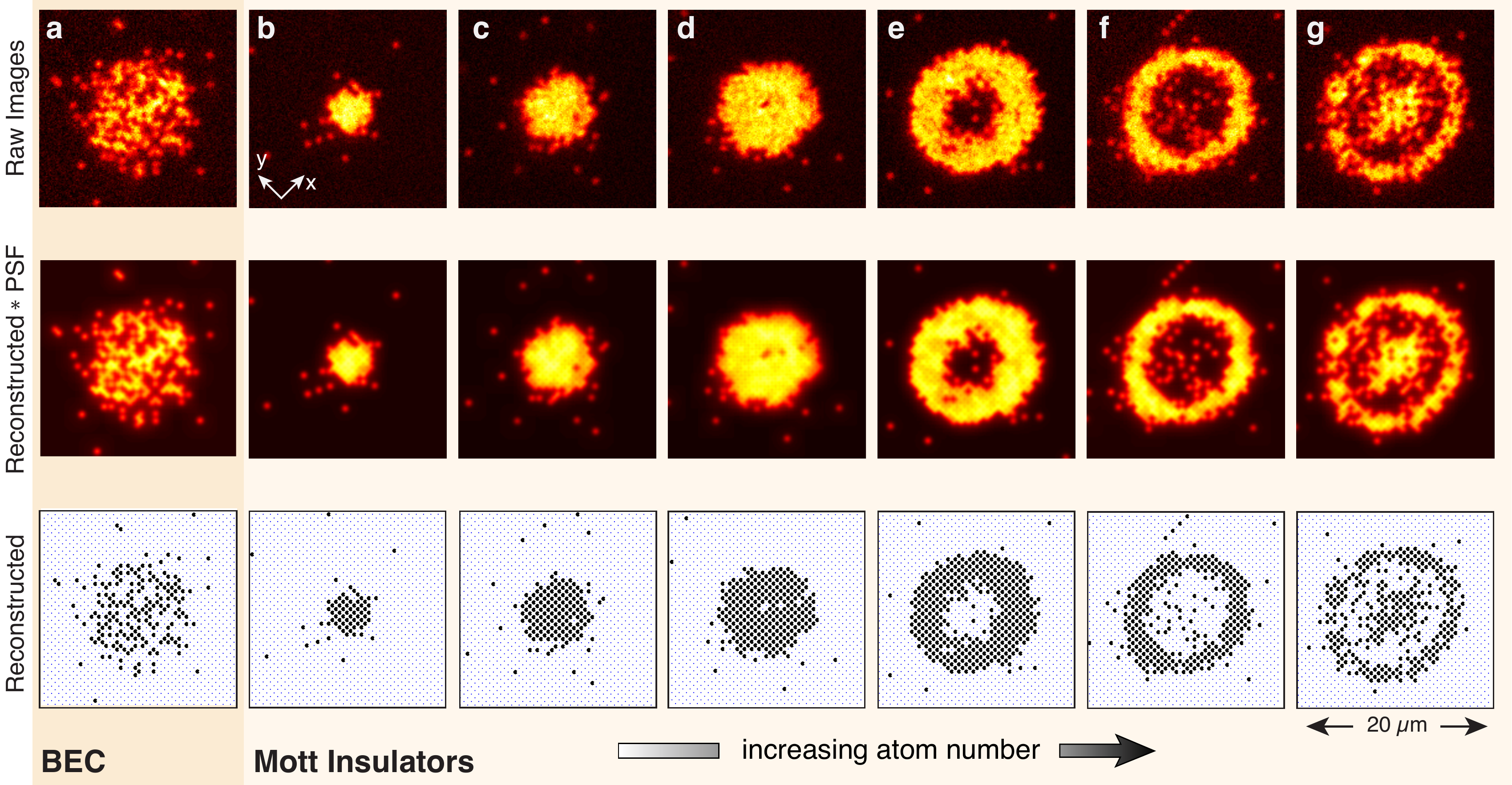}
    \end{center}
    \caption{{\bf High resolution fluorescence images of a BEC and Mott insulators.} Top row: Experimentally obtained images of a BEC {\bf (a)} and Mott insulators for increasing particle numbers {\bf (b-g)} in the zero-tunneling limit. Middle row: Numerically reconstructed atom distribution on the lattice. The images were convoluted with the point spread function of our imaging system for comparison with the original images. Bottom row: Reconstructed atom number distribution. Each circle indicates a single atom, the points mark the lattice sites. BEC and MIs were prepared with  the same in-plane harmonic confinement (see Supplementary Information for the Bose-Hubbard model parameters of our system).\label{fig:bec_mi_series}}
\end{figure*}

Ultracold atoms in optical lattices have proven to be powerful
simulators for the investigation of the static quantum phases and
dynamical evolutions of strongly correlated quantum many-body
systems. Prominent examples include the quantum phase transition
from a superfluid to a Mott insulator
\cite{Fisher:1989,Jaksch:1998,Greiner:2002a,Spielman:2007} and the
fermionized Tonks-Girardeau gas for bosonic particles
\cite{Paredes:2004,Kinoshita:2004}, as well as the recently realized
fermionic Mott insulator \cite{Joerdens:2008a,Schneider:2008a}. In
all these cases, the strong interactions between the particles
compared to their kinetic energy generate intriguing highly
correlated quantum states that are of fundamental interest in
condensed matter physics and promising for practical applications in
quantum information science. For many of these applications it is of
crucial importance to image the correlated many-body systems with
single-atom and single-site resolution. One could then e.g. probe
the evolution from a Poissonian atom number distribution into highly
number squeezed Fock states on a lattice not only globally
\cite{Gerbier:2006a} but also on a local scale. Furthermore one
should be able to directly observe critical phenomena in the in-situ
density or in spin-resolved images of the particles when approaching
a quantum critical point. For applications in quantum information
science, it is essential to address and manipulate single atoms on
individual lattice sites. A prominent example is the one-way quantum
computer \cite{Raussendorf:2001} where local single particle
measurements and operations are carried out after a successful
global entanglement operation that creates a highly-correlated
cluster state.

Over the past years tremendous progress has been made in the
high-resolution and single-atom sensitive detection of atoms on a
lattice
\cite{Nelson:2007,Gericke:2008,Gemelke:2009,Karski:2009,Bakr:2009}.
However, it has only now become possible to apply these techniques
to the detection of strongly correlated quantum systems, in the work
presented here and the work of Bakr et al. \cite{Bakr:2010}.  Here we
report on in-situ fluorescence imaging of a Mott insulator (MI) with
single-atom and single-site resolution. From a {\it single} image,
we reconstruct the atom distribution on the lattice and individual
thermal excitations of the MI become directly visible. This allows
us to observe the number squeezing and the quality of an atomic Mott
insulator down to a single lattice site. Using a simple model
\cite{Gerbier:2007b,Ho:2007} we characterize the average density
distribution and number fluctuations of the quantum system, and use
this for an in-situ temperature measurement. We find excellent
agreement between our theory that assumes global thermal
equilibrium. Furthermore, we show how the incompressible Mott phase
evolves into a compressible normal phase as the temperature is
increased.

%BEGIN EXPERIMENT DESCRIPTION
Our experiments start with an almost pure 2D Bose-Einstein
condensate \cite{Spielman:2007} (BEC) of up to a few thousand
$^{87}$Rb atoms that was prepared  in a single pancake-shaped
antinode of a vertical optical standing wave (beam waist
$w_0=75\,\mu$m) oriented along the $z$-axis (see
Fig.~\ref{fig:schematic}). The lattice depth was  $V_z=26(2)\,E_r$,
where $E_{r}=h^2/(2m \lambda^2)$ is the recoil energy, $m$ denotes
the atomic mass of $^{87}$Rb and $\lambda$ the lattice wavelength.
Additional beams along the $x$- and $y$-directions were used to load
the 2D quantum gas into an optical lattice. All lattice beams had a
wavelength $\lambda=1064$\,nm resulting in a lattice period of
$532$\,nm. We detected the atoms in the lattice by high-resolution
fluorescence imaging through a specially designed microscope
objective with a numerical aperture of $\mbox{NA}=0.68$ and an
optical resolution (FWHM) of $\approx700$~nm at a wavelength of
$780$~nm. For detection, the lattice depths along all three
directions were increased to $V_{x,y,z}/k_B \approx  300\,\mu$K
before an optical molasses induces fluorescence and simultaneously
laser cooled the atoms \cite{Nelson:2007,Bakr:2009} (see Methods).
In the low density thermal clouds (inset in
Fig.~\ref{fig:schematic}) individual atoms are directly visible
above an almost indiscernible background and their positions have a
discrete spacing given by our lattice period (see Supplementary
Information). During the imaging, atom pairs on a lattice site are
immediately lost due to inelastic light-induced collisions
\cite{DePue:1999}. We therefore only detect the particle number
modulo two on each lattice site. This essentially amounts to
recording the parity of the atom number.

The 2D lattice gases used in our experiments are well described by
the Bose-Hubbard model, where particles are restricted to occupy the
lowest energy band of the lattice and their kinetic energy is
characterized by a tunneling matrix element $J$ and an on-site
two-particle interaction energy $U$ (see
Refs.\,\cite{Jaksch:2005,Bloch:2008c}). For a BEC loaded into a weak
lattice potential, $U/J \ll 1$, one expects a Poissonian atom number
distribution on a lattice site $i$, as the classical coherent matter
wave field of a BEC is characterized by Glauber's coherent states.
Such states with an average filling of $\nbari$ per lattice site,
exhibit a corresponding variance in the particle number
$\sigma_i^2=\nbari$. When the interactions between the particles
relative to their kinetic energy are increased, the system undergoes
a quantum phase transition to a Mott insulating state
\cite{Fisher:1989,Jaksch:1998,Greiner:2002a}. For homogeneous
conditions and a 2D simple square lattice, this transition is
expected to occur at $(U/J)_c \simeq 16.4$ (see
Ref.\,\cite{Krauth:1991}), where small shifts of this critical value
have been reported when the system is additionally exposed to an
underlying harmonic trapping potential \cite{Rigol:2009}. In our
case such an additional harmonic confinement was caused by the
Gaussian beam profile of our lattice beams ($1/e^2$ waist of
75$\mu$m) and resulted in an in-plane harmonic confinement with
trapping frequencies  $\omega_x/(2\pi)=72(4)$\,Hz and
$\omega_y/(2\pi)=83(4)$\,Hz for lattice depths of
$V_{x,y}=23(2)\,E_r$. For  $U/J \gg (U/J)_c$ the atomic MI can be
described by neglecting the tunneling energy of the system in the so
called zero-tunneling approximation. The in-trap density
distribution then exhibits a pronounced shell structure of
incompressible regions where the density is pinned to integer values
and increases in a step-like manner from the outer wings to the
inner core
\cite{Jaksch:1998,Foelling:2006,Campbell:2006,Gemelke:2009}. At zero
temperature the particle number variance at a lattice site is then
expected to vanish ($\sigma_i^2=0$) resulting in perfect Fock
states. For low, but still finite temperatures $k_B T \ll U$,
thermal fluctuations can be induced. These fluctuations limit the
quality of the number squeezing and eventually lead to a complete
melting of the characteristic shell structure of a MI when the
temperature is increased above $T_{m} \simeq 0.2 U/k_B$ (see
Refs.~\cite{Gerbier:2007b,Ho:2007}).

%%%%%%%%%%%%%%%%%%%%%%%%%%%%%%%%%%%%%%%%%%%%%%%%%%%%%%%%%%%%%%%%%%%%%%%%%%
%FIG 2
%%%%%%%%%%%%%%%%%%%%%%%%%%%%%%%%%%%%%%%%%%%%%%%%%%%%%%%%%%%%%%%%%%%%%%%%%%

\begin{figure}[!t]
    \begin{center}
        \includegraphics[width=0.75\columnwidth]{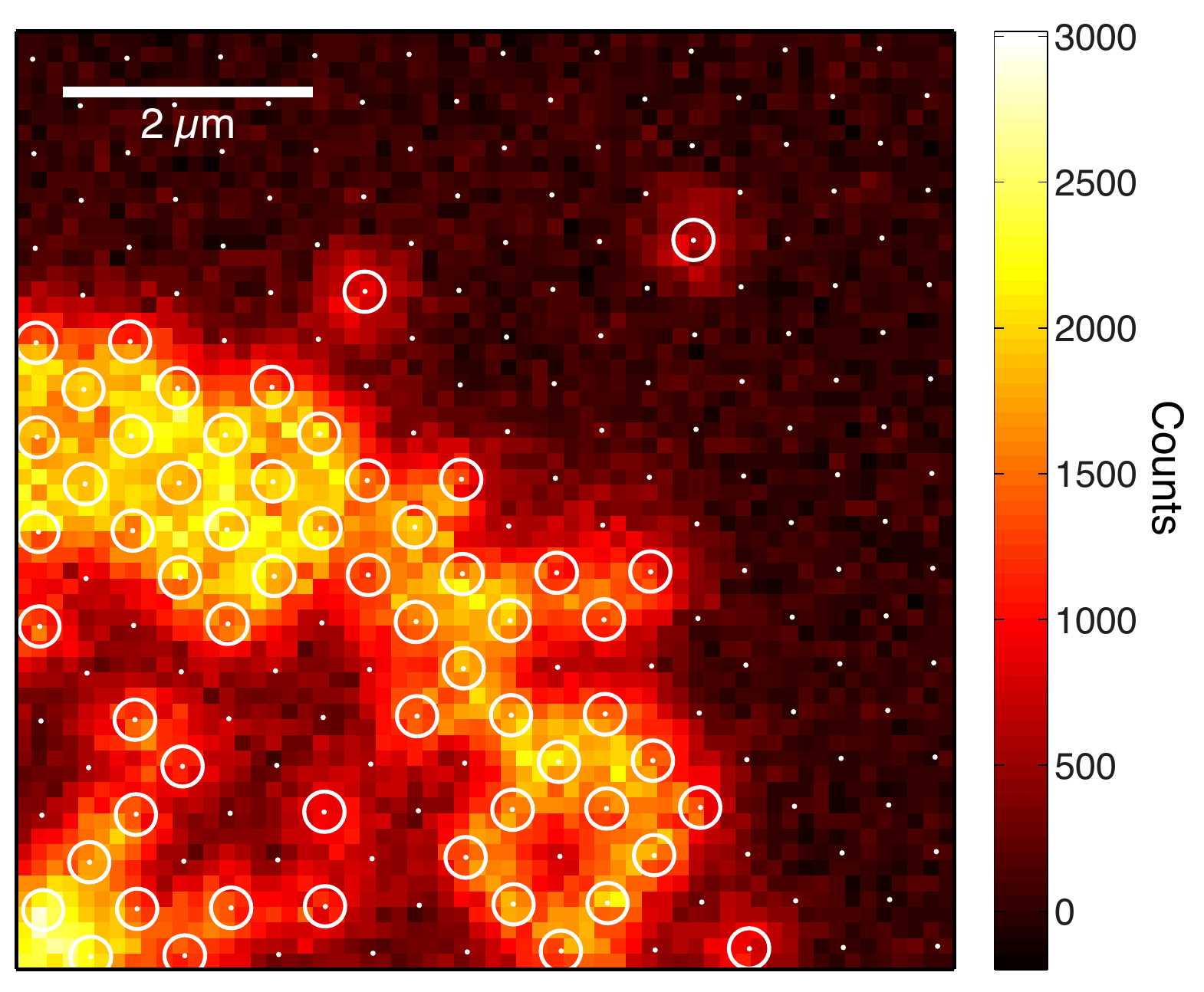}
    \end{center}
    \caption{{\bf Identification of single atoms in a high-resolution image.} The points mark the centers of the lattice sites, circles indicate those sites where our deconvolution algorithm determined the presence of an atom. The image is a zoom into the upper right part of Fig.~\ref{fig:bec_mi_series}g.\label{fig:reconstruction}}
\end{figure}

We monitored the dramatic differences in the density profiles and
the on-site number fluctuations  by imaging the in-trap atom
distributions of a BEC and a MI in the zero-tunneling limit for
different atom numbers and temperatures (see
Fig.~\ref{fig:bec_mi_series}, first row). For the MIs, the lattices
along the $x$- and $y$-directions were increased in $s$-shaped ramps
within 75\,ms up to values of $V_{x,y}=23(2)\,E_r$. To freeze out
the atom distribution of a BEC, we ramped up the lattices within
0.1\,ms. Using the point spread function (PSF) of our optical
imaging system we were able to reconstruct the atom number
distribution on the lattice with single-site and single-atom
resolution via an image processing algorithm (see Methods). It works
reliably even in the regions of high atomic density, as illustrated
in Fig.~\ref{fig:reconstruction}.

To compare the digitally reconstructed atom distribution (see
Fig.~\ref{fig:bec_mi_series}, bottom row) with the raw images, we
show the reconstructed distribution convoluted with the PSF in the
center row of Fig.~\ref{fig:bec_mi_series}. For a BEC with a
Poissonian atom number distribution the average filling one detects
due to the parity measurement in the fluorescence imaging is
$\nbaridet=1/2\left[1-\exp{(-2\nbari)}\right]$, which saturates at
$\nbaridet =0.5$ for $\nbari \gtrsim 1.5$. In this limit, the
detected atom number variance then saturates accordingly at
$\sigma^2_{i,{\rm det}} = 0.25$. Indeed for a BEC, we observed that
the recorded atomic density exhibits large atom number fluctuations
from site to site. In contrast, for a MI we obtain plateaus of
constant integer density, with almost vanishing fluctuations. For
increasing particle numbers, the images in
Fig.~\ref{fig:bec_mi_series} show how successive MI shells are
formed,  which appear as alternating rings of one and zero atoms per
site due to our parity measurement. In both the raw images and the
reconstructed ones, individual defects are directly visible. The
high symmetry of our atom clouds reflects the high optical quality
of our lattice potentials. A small ellipticity is caused by the
different harmonic trapping frequencies $\omega_x$ and $\omega_y$.

%%%%%%%%%%%%%%%%%%%%%%%%%%%%%%%%%%%%%%%%%%%%%%%%%%%%%%%%%%%%%%%%%%%%%%%%%%
%RECONSTRUCTION + TEMPERATURE MEASUREMENT
%%%%%%%%%%%%%%%%%%%%%%%%%%%%%%%%%%%%%%%%%%%%%%%%%%%%%%%%%%%%%%%%%%%%%%%%%%

\begin{figure}[!t]
    \begin{center}
        \includegraphics[width=0.95\columnwidth]{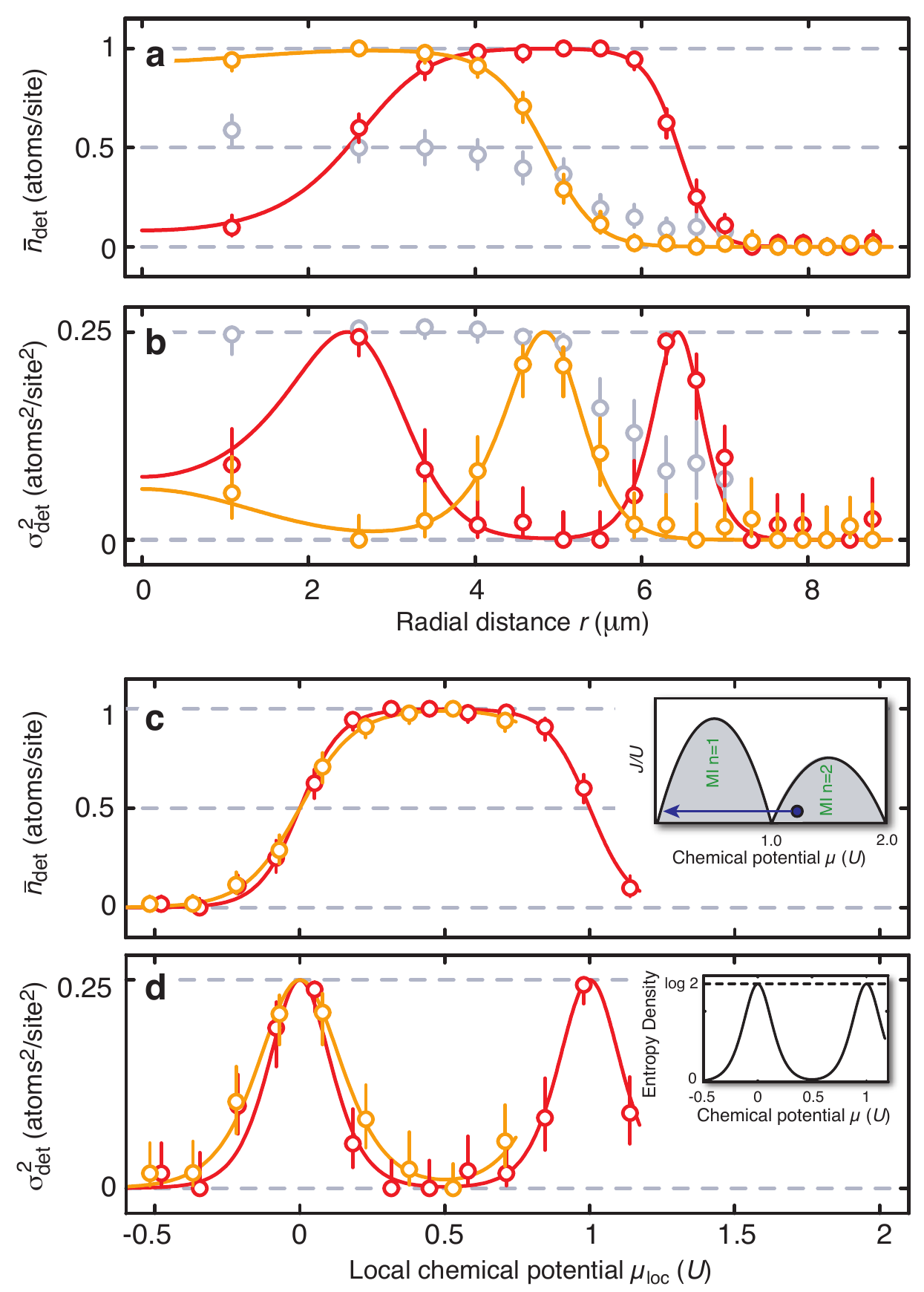}
    \end{center}
    \caption{{{\bf Radial atom density and variance profiles.} Radial profiles were obtained from the digitized reconstructed images by azimuthal averaging. {\bf a}, {\bf b}, Yellow and red points correspond to the  $n=1$ and $n=2$ MI images of Fig.\,\ref{fig:bec_mi_series}d,e. The grey points were obtained from a BEC (data from Fig.\,\ref{fig:bec_mi_series}a) for reference. The displayed statistical errors are Clopper-Pearson 68\% confidence intervals for the binomially distributed number of excitations. For the MIs both density and variance profiles are fitted simultaneously with the model functions of Eqs.~(\ref{eq:nBar}) and (\ref{eq:variance}) (see Methods) with $T$, $\mu$ and $r_0^2=2 U/(m\omega_x\omega_y)$ as free parameters. For the two curves, the fits yielded temperatures $T=0.090(5)U/k_B$ and $T=0.074(5)U/k_B$, chemical potentials $\mu=0.73(3)U$ and $\mu=1.17(1)U$, and radii $r_0=5.7(1)\,\mu$m and $r_0=5.95(4)\,\mu$m respectively. From the fitted values of $T$, $\mu$ and $r_0$, we determined the atom numbers of the system to $N=300(20)$ and $N=610(20)$.
 {\bf c}, {\bf d}, The same data plotted versus the local chemical potential using the local-density approximation. The inset of {\bf c} is a Bose-Hubbard phase diagram ($T=0$) showing the transition between the characteristic MI lobes and the superfluid region. The line starting at the maximum chemical potential $\mu$ shows the part of the phase diagram existing simultaneously at different radii in the trap due to the external harmonic confinement. The inset of {\bf d} is the entropy density calculated for the displayed $n=2$ MI.}\label{fig:radial_average}}
\end{figure}

We used the reconstructed site occupation numbers to determine the
temperature of the sample based on a single image. For deep
lattices, $U/J\simeq 300$, as used in our experiments for MIs,
tunneling becomes completely suppressed such that coherent
particle-hole fluctuations are expected to be negligible and defects
are only induced by thermal fluctuations. The symmetry of our clouds
allowed us to average the data azimuthally, taking into account the
ellipticity, and to obtain radial profiles for the average density
$\bar n_{\rm det}(r)$ and variance $\sigma^2_{\rm det}(r)$ (see
Fig.~\ref{fig:radial_average}a,b and Methods). We fitted analytic
expressions derived in the zero-tunneling regime (see Methods) to
our data. The results of such a fit for an $n=1$ ($0<\mu/U<1$) and
an $n=2$ ($1<\mu/U<2$) MI are displayed in
Fig.~\ref{fig:radial_average}a,b. The MI regions can be identified
as connected regions of constant integer density and vanishing
on-site number fluctuations, which in the atomic limit of the
Hubbard model signify the presence of incompressible Mott domains
\cite{Rigol:2009}. For both density profiles and atom number
variances we find excellent agreement between the experimental data
and the theoretical model for all radial distances. This supports
the assumption that our system is in global thermal equilibrium, in
contrast to Ref.~\cite{Hung:2010}. The extracted temperatures of
$T=0.090(5)U/k_B$ and $T=0.074(5)U/k_B$ for the $n=1$ and $n=2$ data
are well below the MI melting temperature $T_{m}$. Our
temperature estimates are conservative since all defects are
attributed to thermal excitations in our model. However, defects
might also arise due to ``collateral damage'' caused by atoms
undergoing the light-induced collisions. For reference, we show the
corresponding data obtained by freezing out the atom distribution of
a BEC. We observe the expected saturation of $\bar n_{\rm det}$ at
0.5 together with a maximum variance of $\sigma_{\rm det}^2$ at
0.25. We note that the thermal shells surrounding a MI core also
exhibit this maximum variance and can be as narrow as 1-2 lattice
sites. In Fig.~\ref{fig:radial_average}c,d we plot both MI data sets
versus local chemical potential. In a single image, we thus mapped
out an entire line in the phase diagram as illustrated in the inset
of Fig.~\ref{fig:radial_average}c. The slightly different
temperatures of the two MIs are clearly visible in the different
widths of the variance curves.

Our measurements also confirm with unprecedented clarity that the
entropy of the strongly correlated quantum gas is concentrated
around the Mott insulating regions, whereas in the center of a MI,
for local chemical potentials of $\mul=(n+1/2)U$, number
fluctuations are completely suppressed and the entropy density is
essentially zero.  For the lowest observed temperature of
$T=0.074(5)U/k_B$  we calculate a 99.7(1)\% probability for unity
occupation in the center of the $n=1$ Mott-insulating plateau. Using
the zero-tunneling model we can also extract the total entropy per
particle for our system $S/(Nk_B)=0.34(2)$ (see Methods) which is
around the critical entropy for quantum magnetism
\cite{Capogrosso-Sansone:2010}.

%%%%%%%%%%%%%%%%%%%%%%%%%%%%%%%%%%%%%%%%%%%%%%%%%%%%%%%%%%%%%%%%%%%%%%%%%%
% MELTING OF A MI
%%%%%%%%%%%%%%%%%%%%%%%%%%%%%%%%%%%%%%%%%%%%%%%%%%%%%%%%%%%%%%%%%%%%%%%%%%
\begin{figure}[!h]
    \begin{center}
        \includegraphics[width=0.8\columnwidth]{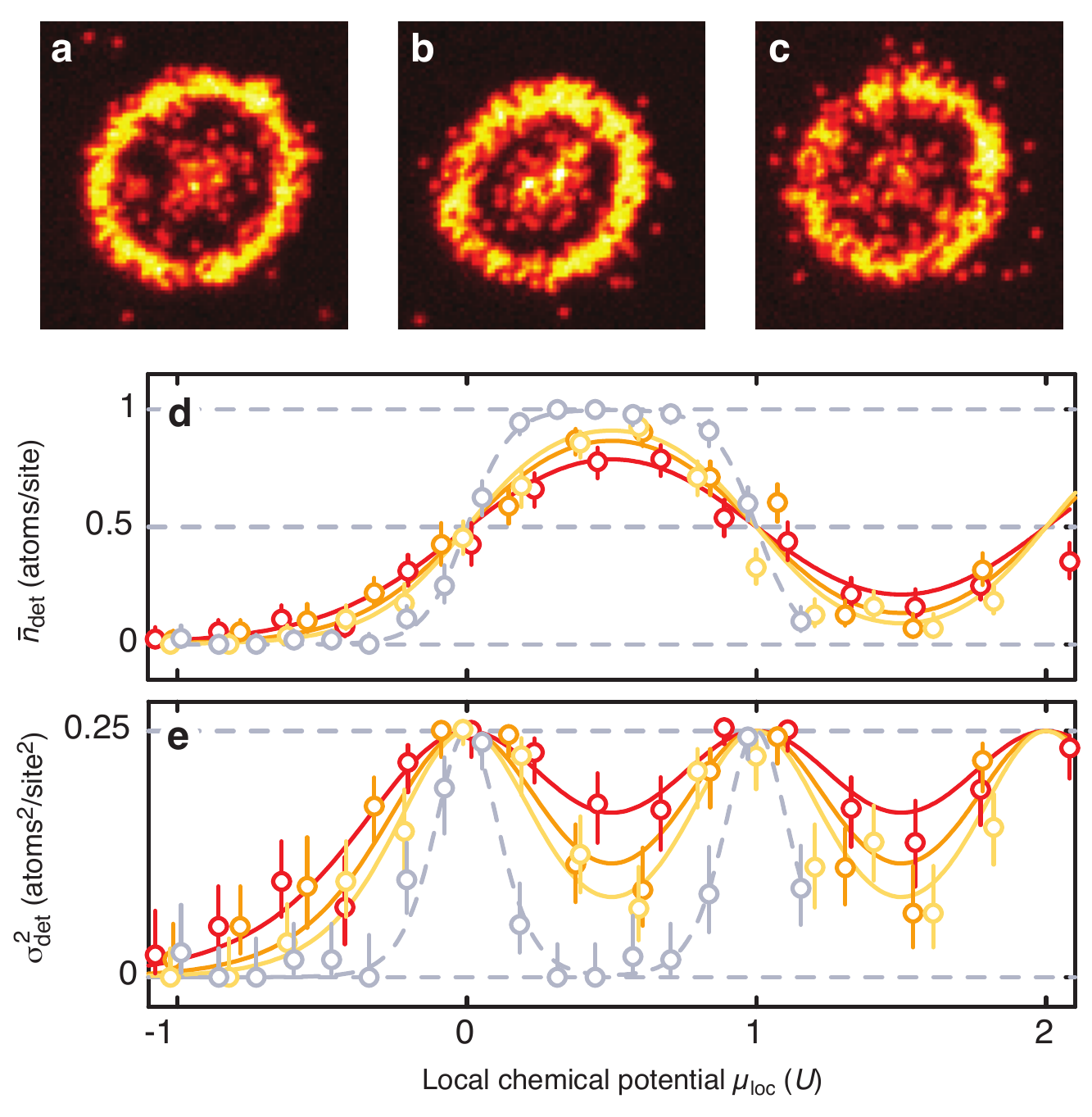}
    \end{center}
    \caption{{\bf Melting of a Mott Insulator.}
{\bf a}-{\bf c} Strongly correlated atomic samples for three
different temperatures, and constant total chemical potentials in
the zero-tunneling limit. For higher temperatures, an increased
number of independent particles or holes appear. Data shown was
binned 2x2. {\bf e, f}, Density and variance profiles as a function
of chemical potential, determined as described in the caption of
Fig.\,\ref{fig:radial_average}. Red, orange and yellow points
correspond to the datasets from {\bf a}, {\bf b}, and {\bf c},
respectively. Light blue points correspond to the low-temperature
$n=2$ MI of Fig.\,\ref{fig:bec_mi_series}e and
Fig.\,\ref{fig:radial_average} with $T=0.074(5)U/k_B$ and
$\mu=1.17(1)U$. The parameters extracted from the radial fits are
$T=0.17(1)U/k_B$, $\mu=2.08(4)U/k_B$, for {\bf a}, $T=0.20(2)U/k_B$,
$\mu=2.10(5)U/k_B$, for {\bf b} and $T=0.25(2)U/k_B$,
$\mu=2.06(7)U/k_B$, for {\bf c}. \label{fig:hot_mi} }
\end{figure}

Finally, we show how a Mott insulator melts, as the temperature (or
entropy) of the quantum gas is increased (see
Fig.\,\ref{fig:hot_mi}). At constant total chemical potential but
increasing temperatures, one observes that the Mott domains
gradually vanish. Although there is no sharp transition to a normal
fluid state in this case, Mott plateaus and number squeezing degrade
rapidly, once  $T\gtrsim T_m$ (see Figs.\,\ref{fig:hot_mi}d,e), as
predicted in Ref.\,\cite{Gerbier:2007b}.

%%%%%%%%%%%%%%%%%%%%%%%%%%%%%%%%%%%%%%%%%%%%%%%%%%%%%%%%%%%%%%%%%%%%%%%%%%
%OUTLOOK
%%%%%%%%%%%%%%%%%%%%%%%%%%%%%%%%%%%%%%%%%%%%%%%%%%%%%%%%%%%%%%%%%%%%%%%%%%
In summary, we have demonstrated single-site {and single-particle}
resolved detection of a strongly correlated system in an optical
lattice. Our method can be extended to investigate quantum critical
phenomena, density-density correlations or even non-local
string-operators that are inaccessible in condensed matter
experiments. Our imaging system can be used to focus an off-resonant
laser beam onto a single atom and thereby allow single-site
manipulation of the atomic qubits \cite{Zhang:2006}. This opens a
new avenue for experiments with ultracold quantum gases, where e.g.
novel cooling schemes may be applied by accessing regions of high
entropy \cite{Bernier:2009}. For future work it would be interesting
to investigate how entropy propagates in strongly correlated
systems, after injecting it on a local scale into the system. Atoms
in the MI with one atom per lattice site are also very promising as
a quantum register for scalable quantum computing, especially with
the very low defect density shown in this work.

%%%%%%%%%%%%%%%%%%%%%%%%%%%%%%%%%%%%%%%%%%%%%%%%%%%%%%%%%%%%%%%%%%%%%%%%%%
%METHODS SUMMARY
%%%%%%%%%%%%%%%%%%%%%%%%%%%%%%%%%%%%%%%%%%%%%%%%%%%%%%%%%%%%%%%%%%%%%%%%%%

We would like to thank Rosa Gl\"ockner and Ralf Labouvie for
assistance during the setup of the experiment and Stefan Trotzky for
helpful discussions. We acknowledge funding by MPG, DFG, Stiftung
Rheinland-Pfalz f\"{u}r Innovation, Carl-Zeiss Stiftung, EU
(NAMEQUAM, AQUTE, Marie Curie Fellowships to J.F.S. and
    M.C.).

%%%%%%%%%%%%%%%%%%%%%%%%%%%%%%%%%%%%%%%%%%%%%%%%%%%%%%%%%%%%%%%%%%%%%%%%%%
%METHODS
%%%%%%%%%%%%%%%%%%%%%%%%%%%%%%%%%%%%%%%%%%%%%%%%%%%%%%%%%%%%%%%%%%%%%%%%%%
\section*{Methods}

{\bf Preparation of a single 2D system}. We started by loading atoms
from a magnetic quadrupole trap into a single beam optical dipole
trap (1064\,nm, beam waist $w_0=40\,\mu$m). By translating the focus
of the dipole trap using mirrors on a motorized micrometer stage,
the atoms were transported in front of the high resolution imaging
system within $2.5$~s. A magnetic quadrupole field whose center is
shifted below the trap compressed the cloud in the axial direction
of the dipole trap laser beam. After 500\,ms of evaporative cooling
in this hybrid trap configuration, we transferred the atoms into the
$z$-lattice. It is oriented along the optical axis of the imaging
system, parallel to gravity, and superimposed with the hybrid trap.
Initially we populated 60 antinodes (slices) of the standing wave,
creating independent 2D systems. To extract a single 2D system, we
used position dependent microwave transfer in a magnetic field
gradient. The gradient with $\partial B/\partial z = 24$\,G/cm
(together with a bias field of 32~G) was produced with a single coil
placed 50\,mm above the atoms; the coil axis coincides with the
$z$-lattice beam. Our magnetic field gradient results in a position
dependent frequency shift $\partial \nu/\partial z \sim
5$~kHz/$\mu$m of the $\ket{F=1, m_F=-1}$ $\leftrightarrow$
$\ket{F=2, m_F=-2}$ transition. An initial microwave frequency sweep
over 1\,MHz brought all atoms from $\ket{F=1, m_F=-1}$ to $\ket{F=2,
m_F=-2}$. We then transferred atoms from one slice back to the
$\ket{F=1, m_F=-1}$ state using a resonant Blackman pulse of 5~ms
duration. All atoms remaining in $F=2$ were removed from the trap by
a laser pulse resonant with the $F=2\rightarrow F'=3$ transition. We
then evaporatively cooled the atoms by ramping down the intensity of
the $z$-lattice from $300\,E_r$ to $\sim25\,E_r$ within 1.5~s, while
simultaneously tilting the potential along the horizontal direction
with a magnetic field gradient \cite{Hung:2008}. Depending on the
end point of this evaporation, we created BECs with atom numbers
ranging from $50-2000$.

{\bf Imaging single atoms in the lattice.} Our microscope objective
was custom made (Leica Microsystems) and is located outside of the
vacuum chamber with a working distance of 13\,mm. We detected the
atoms by illuminating them with an optical molasses, red detuned
with respect to the free space resonance by 45\,MHz. It consists of
two pairs of retroreflected laser beams superimposed with the
horizontal lattice axes. A third $z$-molasses laser beam aligned in
reverse direction through the imaging system provided cooling in the
vertical direction. The total scattering rate from all laser beams
was $\sim60$\,kHz. With our total detection efficiency of $\sim9~\%$
(solid angle 15\%, transmission of all optical elements 71\%, camera
quantum efficiency 85\%), we collect about 5000\,photons/atom within
our illumination time of 900\,ms. To ensure a homogeneous
illumination of all atoms, we scanned the retro-reflecting mirrors
of the horizontal molasses beams (at a frequency of 300~Hz and
400~Hz) using piezo elements and mutually detuned the molasses beams
by 43\,Hz. Additionally, we spatially scanned the $z$-molasses beam
across the cloud at a frequency of 100\,Hz and an amplitude of
35\,$\mu$m. We also corrected for an etaloning effect of the CCD
camera, which caused a spatially dependent signal strength. We
carefully optimized the molasses parameters to minimize hopping of
the atoms to adjacent lattice sites, by taking two consecutive
images of the same cloud for the first 400 ms and the last 400 ms of
our 900 ms illumination period. From the analysis of several of
these double images, we found that about 0.5\% of the atoms hop
during the molasses illumination. Before switching on the molasses,
we removed the atoms in the doubly occupied sites with a 50\,ms
pulse on the $F=2$ to $F'=3$ transition, which is 6.8\,GHz red
detuned for the atoms in $F=1$, but efficiently excites into the
molecular potentials.

For the horizontal optical lattice laser beams we used two single
mode fiber amplifiers seeded with the same narrowband solid-state
laser, whereas the vertical lattice beam was derived from an
independent solid-state laser. The horizontal axes were mutually
detuned by 220~MHz and had orthogonal polarizations. We obtained
about 10\,W per lattice axis at the experiment, yielding trap depths
of about 300~$\mu$K.

{\bf Reconstruction of the atom number distribution.} We developed a
deconvolution algorithm to reconstruct the atom number distribution
from a fluorescence image. It uses a model of the point spread
function (PSF) of our imaging system that was determined from
averaging over many images of isolated individual atoms (see
Supplementary Information). The algorithm tries different
model-configurations for each lattice site and its nearest neighbors
in order to minimize the difference of the original image with the
reconstructed one. This reconstructed image was obtained by
convoluting the atom number distribution with our PSF (see centre
row of Fig.~\ref{fig:bec_mi_series}). The algorithm allows for a
variance of the fluorescence level of each atom within $\pm 20\%$ of
the mean photon counts. These varying fluorescence levels partially
arise from the inhomogeneous intensity of the molasses light. We
additionally found an increased fluorescence level of about
10\%-20\% in the center of very dense $n=1$ shells of a MI, compared
with the isolated atoms in the outer part of the images. This effect
might arise from the partial coherence of the light scattered by the
atoms, combined with their regular distribution in the lattice.

We have evaluated the fidelity of the reconstruction algorithm by
creating simulated images of a known atom distribution using the PSF
of our imaging system, the poissonian and superpoissonian noise
contributions of the light hitting the EMCCD camera (including the
amplification process), and the site-to-site fluorescence
fluctuations of $\pm20\%$. Running the reconstruction algorithm over
several hundred of such randomly generated images of MIs at finite
temperatures, we find a reconstruction fidelity of $\sim99.5\%$. In
our experiment, the main limitations of the fidelity are atom losses
during the detection process due to collisions with background gas
atoms. We measured that about 1\% of the atoms are lost during the
900\,ms detection period, which corresponds to a trap lifetime of
$\sim75$\,s.

{\bf Radial atom number distribution and variance.} In the
zero-tunneling regime (the ``atomic limit" of a MI), the atom number
distribution at a lattice site at radius $r$ is given by
$P_r(n)=e^{\beta[\mul(r) n - E_n]}/Z(r)$, where
$Z(r)=\sum_ne^{\beta[\mul(r) n - E_n]}$ is the grand canonical
partition function, $\beta=1/(k_B T)$, $\mul(r)$  the local chemical
potential and $E_n=U n(n-1)/2$ is the interaction energy. Using a
local density approximation (LDA), we define $\mul$ in terms of the
global chemical potential $\mu$ and the external harmonic trapping
confinement: $\mul(r)=\mu-\frac{1}{2}m(\omega^2_xx^2+\omega^2_y
y^2)$. Taking the light induced losses into account we calculate the
expected detected density at different radii:
\begin{equation}
    \label{eq:nBar} \bar n_{\rm det}(r) =\frac{1}{Z(r)}\sum_n \text{mod}_2(n) e^{\beta[\mul(r) n - E_n]}
\end{equation}
In the presence of light induced collisions $\overline{n_{\rm
det}^2}(r)=\bar n_{\rm det}(r)$ and the detected variance is
therefore simply
\begin{equation}
    \label{eq:variance} \sigma^2_{\rm det}(r)=\bar n_{\rm det}(r)-\bar n_{\rm det}^2(r).
\end{equation}
We extracted radial density and variance profiles from the
reconstructed two-dimensional atom distribution of a single image.
For this, we first determined the center of the cloud, and then
binned the lattice sites according to their distance from the
center, thereby correcting for the ellipticity of 10\%. The bin
sizes were chosen larger near the center  to have sufficient
statistics.

We fitted the experimental profiles to Eqns. (\ref{eq:nBar}) and
(\ref{eq:variance}) and  extracted the temperature and the global
chemical potential. These can then be used to calculate the original
atom number distribution $P_r(n)$. Inserting the radius
corresponding to $\mul=0.5$ we extract the maximal theoretical unity
occupation probability. We can furthermore calculate the local
entropy density $S_\mathrm{loc}(r)=-k_B \sum_n P_r(n)
\mathrm{ln}[P_r(n)]$.  Summing the density and entropy density over
the lattice sites we calculate the total number of particles
$N=591(9)$ and the total entropy $S/k_B=200(13)$ given the fitted
values of $T$ and $\mu$ from the $n=2$ data of Fig.
\ref{fig:radial_average}. This gives the entropy per particle $S/(N
k_B)=0.34(2)$.

\bibliographystyle{naturemag}
\bibliography{Preprint_Sherson}

\section*{Supplementary Information}

\begin{figure}[!b]
\begin{center}
\includegraphics[width=0.95\columnwidth]{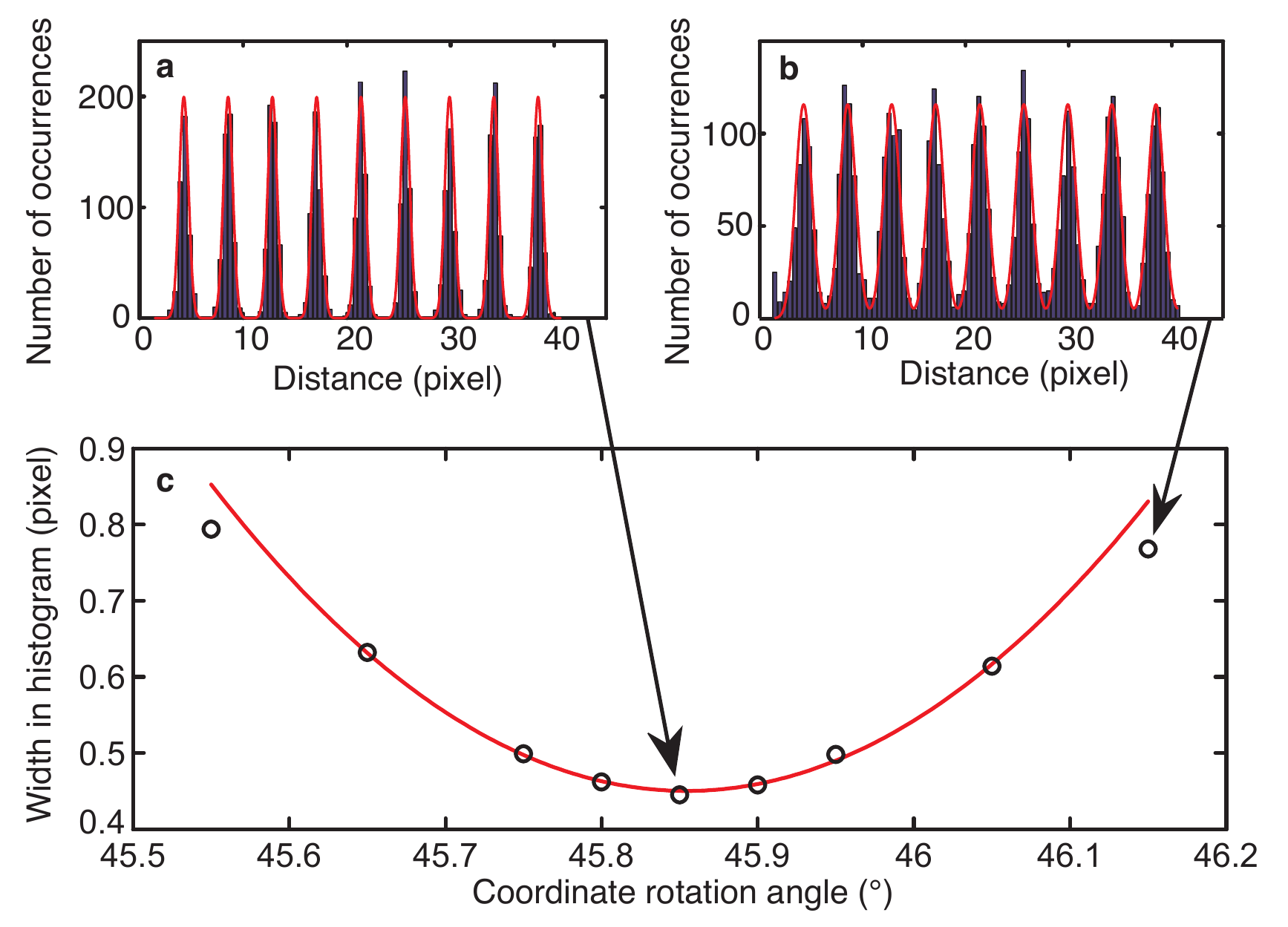}
\end{center}
\caption{{\bf Determination of the lattice angles}. {\bf a,b}
Histogram of the distances between the center positions of
individual atoms projected in a coordinate system rotated by an
angle $\theta$. The line is a fit to a sum of equidistant Gaussians.
{\bf c} The width of the fitted Gaussians show a clear minimum. The
red line is a parabolic fit and yields a minimum rotation angle at
$\theta=45.85(1)^\circ$.  \label{fig:ContrastVsAngle}}
\end{figure}

{\bf Determination of lattice angles and spacing.} To characterize
our imaging system and to determine the lattice structure, we used a
fluorescence image of a dilute thermal could, similar to the inset
in Fig.~\ref{fig:schematic} of the main text. The lattice axes are
oriented at approximately $\pm45^\circ$ with respect to our images.
A precise determination of this angle and the lattice spacing is
needed so that the deconvolution algorithm works with high fidelity.
We first determined the center positions of isolated atoms from this
image by a simple fitting algorithm. The histogram of the mutual
distances projected in a coordinate system rotated by an angle
$\theta$ clearly shows the periodicity of the lattice (see
Fig.~\ref{fig:ContrastVsAngle}a,b) and the visibility of the pattern
depends very sensitively  on $\theta$. For a quantitative analysis,
we fit a sum of equidistant Gaussians to the histogram. The width of
the Gaussians  for different values of $\theta$
(Fig.~\ref{fig:ContrastVsAngle}c) shows a clear minimum at
$\theta=45.85(1)^\circ$. We obtained a similar graph for the other
lattice axis and found an angle of $-45.55(1)^\circ$. The distance
of the Gaussians is $4.269(4)$ pixel which corresponds to  lattice
period of $532$\,nm. Thus, our magnification factor is 128.4(2) and
one pixel of the CCD camera corresponds to 124.6(1)\,nm in the
object plane. The angles and lattice spacing determined by this
method are used as fixed parameters for our deconvolution algorithm.
We also found that the phases of the two lattice axes slightly drift
from shot to shot. They are determined for each image by fitting the
center positions of single atoms in the outer part of the images.

\begin{figure}[!t]
\begin{center}
\includegraphics[width=0.6\columnwidth]{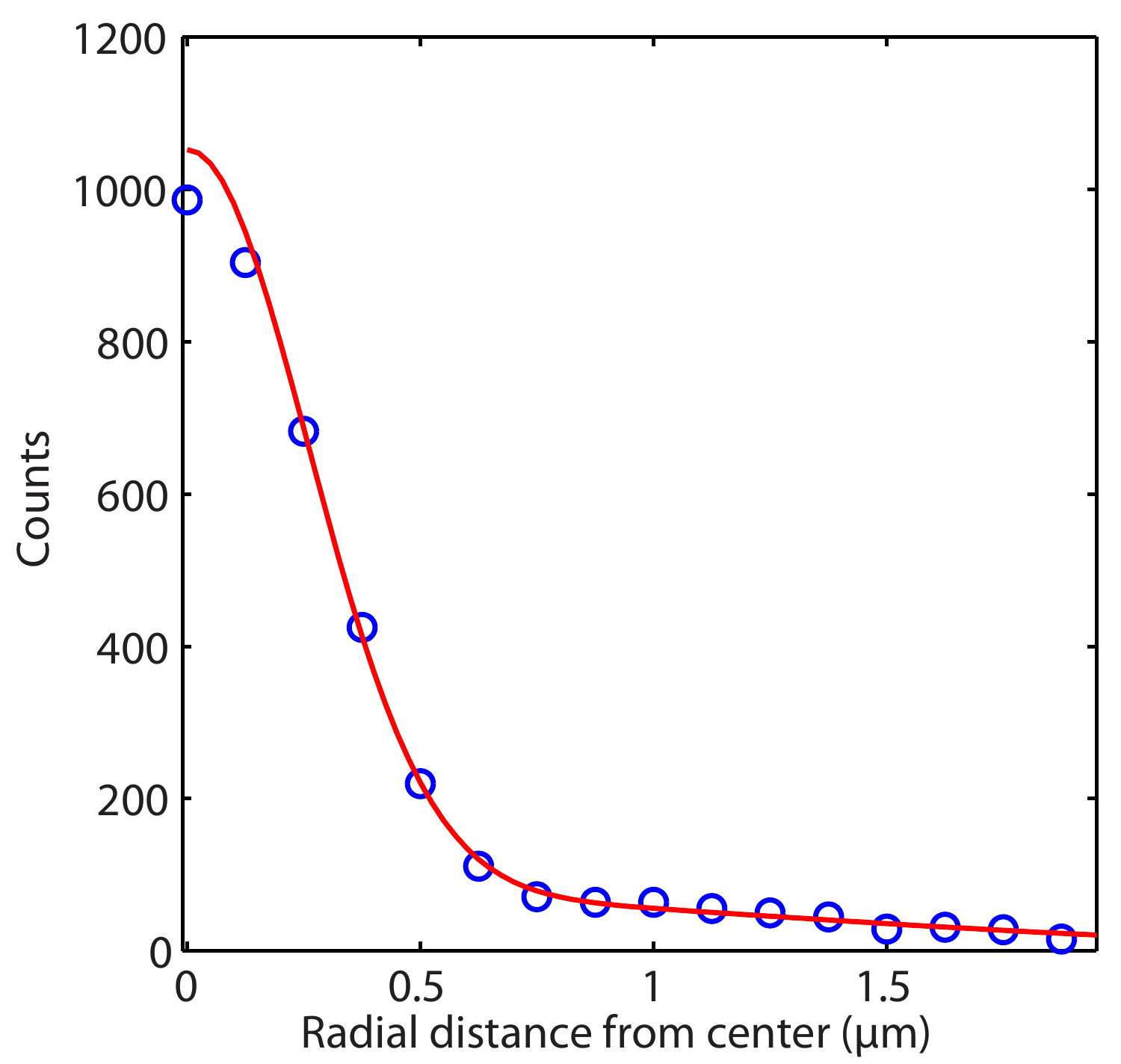}
\end{center}
\caption{{\bf Azimuthal average of our experimentally obtained point spread function.}  The data was obtained by averaging over 68 signals of single atoms. The line is a fit with the double Gaussian of Eq.~(\ref{eq:PSF}) and yields $\sigma_1=2.06(5)$ pixels, $\sigma_2 = 9.6(1.2)$ pixels, $a = 0.075(2)$ and $C=1050(7)$.
\label{fig:PSF}}
\end{figure}

{\bf Determination of the point spread function.}
We determined our point spread function (PSF) from the fluorescence image of a dilute atomic cloud.
We summed the fluorescence image of many individual atoms that were isolated from their neighbors by more that 12 pixels. The summed image is almost radially symmetric and we computed an azimuthal average (see Fig.~\ref{fig:PSF}). We expect our PSF to be a convolution of an Airy disk  with a Gaussian, taking into account fluctuations of the lattice with respect to the imaging system and the width of the atomic wavepacket in the potential wells. Due to this convolution, the first minimum of the airy pattern is not visible in our averaged signal.
We found that our PSF can be well approximated by a double Gaussian:
\begin{eqnarray}\label{eq:PSF}
    PSF(x,y) &= &C \left[(1-a) \exp{\left(-0.5 (x^2+y^2) /
    \sigma_1^2\right)}\right.\nonumber\\
                    & + &\left.a \exp{\left(-0.5 (x^2+y^2) / \sigma_2^2\right)} \right]
\end{eqnarray}
with widths $\sigma_1$, $\sigma_2$ and a parameter $a$ describing
the relative amplitudes. The maximum fluorescence level $C$ varies
from day to day and is in the range of 800-1200 counts. \\

{\bf Single-band Hubbard parameters}

For a $z$-lattice depth of $V_z=26 E_r$, the single-band Hubbard
parameters for our 2D system are given in  the following table:\\

\begin{tabular}{|c|c|c|}
  \hline
  $V_{x,y} (E_r)$ & $J/h$ (Hz) & $U/h$ (Hz)\\
  \hline
  5 & 134 & 382\\
  10 & 39 & 607\\
  15 & 13 & 776\\
  20 & 5  & 917\\
  25 & 2  & 1039\\
  \hline
\end{tabular}\\

Note that one expects the the single-band Hubbard parameters to be
slightly renormalized due to multi-orbital effects.

\end{document}